\documentclass[a4paper]{PoS}

\title{Highlights from the VERITAS AGN Observation Program}

\ShortTitle{Highlights from the VERITAS AGN Observation Program}

\author{\speaker{Wystan Benbow} for the VERITAS Collaboration\thanks{https://veritas.sao.arizona.edu/}\\
        Harvard-Smithsonian Center for Astrophysics, 60 Garden St, Cambridge, MA, 02180, USA\\
        E-mail: \email{wbenbow@cfa.harvard.edu}\\}

\abstract{VERITAS is one of the world's most sensitive detectors of
  astrophysical VHE (Very High Energy; E $>$ 100 GeV) $\gamma$-rays.  
  This array of four 12-m imaging atmospheric-Cherenkov telescopes 
  has operated for $\sim$10 years, and nearly 5000 hours of observations
  have been targeted on active galactic nuclei (AGN).  These studies
  of blazars and radio galaxies have resulted in 36 detections.  Most
  of these detections are accompanied by contemporaneous, broadband 
  observations, which enable detailed studies of the underlying
  jet-powered processes. Recent highlights from the VERITAS AGN
 observation program and scientific results are presented. }

\FullConference{35th International Cosmic Ray Conference --- ICRC2017\\
		10--20 July, 2017\\
		Bexco, Busan, Korea}

\begin{document}

\section{Introduction}

AGN are the most numerous class of identified VHE $\gamma$-ray emitter.
They comprise  $\sim$35\% of the VHE sky catalog with 70 AGN VHE detected
as of {\it ICRC2017} \cite{TEVCAT}.   These luminous objects
emit non-thermal radiation across the entire broadband spectrum and
the VHE band is generally located above the high-energy peak of their
signature double-humped spectral energy distribution (SED).  All VHE
AGN possess jets, and their $\gamma$-ray emission is believed to be
produced in a compact region within these accretion-powered jets, near 
the AGN's central supermassive black hole.  Although four nearby ($z<0.06$) FR-I radio galaxies are VHE
detected, most ($\sim$94\%) VHE AGN are blazars, a class of AGN with jets pointed along the line-of-sight to the
observer.  The VHE blazar population includes four subclasses:  47 high-frequency-peaked BL\,Lac objects (HBLs),
8 intermediate-frequency-peaked BL Lac objects (IBLs), 
2 low-frequency-peaked BL Lac objects (LBLs),
and 6 flat-spectrum radio quasars (FSRQs), as well as 3 objects
whose blazar sub-classifications are uncertain.  The VHE
blazar catalog covers a redshift range from $z = 0.030$ to $z=0.944$,
noting that $\sim$80\% of the objects have either $z<0.3$ or uncertain redshift.
The general proximity of these objects is due to a combination of
energetics requirements and the effects of the extragalactic background
light (EBL) which attenuates VHE photons in an energy- and
distance-dependent manner.

There are two empirical qualities that generally describe VHE AGN:
their observed VHE photon spectra are often soft 
($\Gamma_{obs} \sim 3 - 5$) and their observed VHE flux
is almost always variable. In most cases, the harder the
VHE spectrum is the more interesting the target becomes scientifically, noting
that very few VHE AGN are detected above 1 TeV.  While flux variability
is common, and about a third of VHE AGN are only detected during flares, most VHE AGN 
only show a factor of 2-3 in VHE flux variations with
notable episodes (see, e.g., \cite{PKS2155_flare}) of rapid (minute-scale), large-scale (factor of 100) flux
variations being very rare.  It is worth noting that the observed time
scales for these smaller variations (days to years) often depends
on brightness of the objects in the VHE band, with shorter-duration variations
only seen during isolated flaring episodes or for only the brightest objects.  This could be a sensitivity
effect, as many of the VHE AGN require long integration times for
detection with current instruments.

The VERITAS AGN (radio galaxy and blazar) program focuses on making precision measurements 
of their VHE spectra and variability patterns, while leveraging
contemporaneous multi-wavelength (MWL) observations from both ground-
and space-based facilities.  Its main component is a long-term study of the existing VHE AGN
population in a manner that emphasizes the regular search for, and intense
observation of, major flaring episodes.  Independent of any successful
flare identification, the regular sampling of each VHE AGN aims to build
high-statistics data sets to enable fully-constrained modeling of each
VHE AGN's SED (see, e.g., \cite{1ES0229_paper}).  
The long-term MWL light curves should also allow
for flux and spectral correlation studies that may indicate
commonalities in the origin of each AGN's emission.  VERITAS AGN
studies are also useful for a variety of cosmological measurements and
have been used to to constrain the strength of the intergalactic magnetic field \cite{VERITAS_IGMF} and
the density of the EBL \cite{Elisa_ICRC}.

\begin{figure*}[t]
   \centerline{ {\includegraphics[width=3.9in]{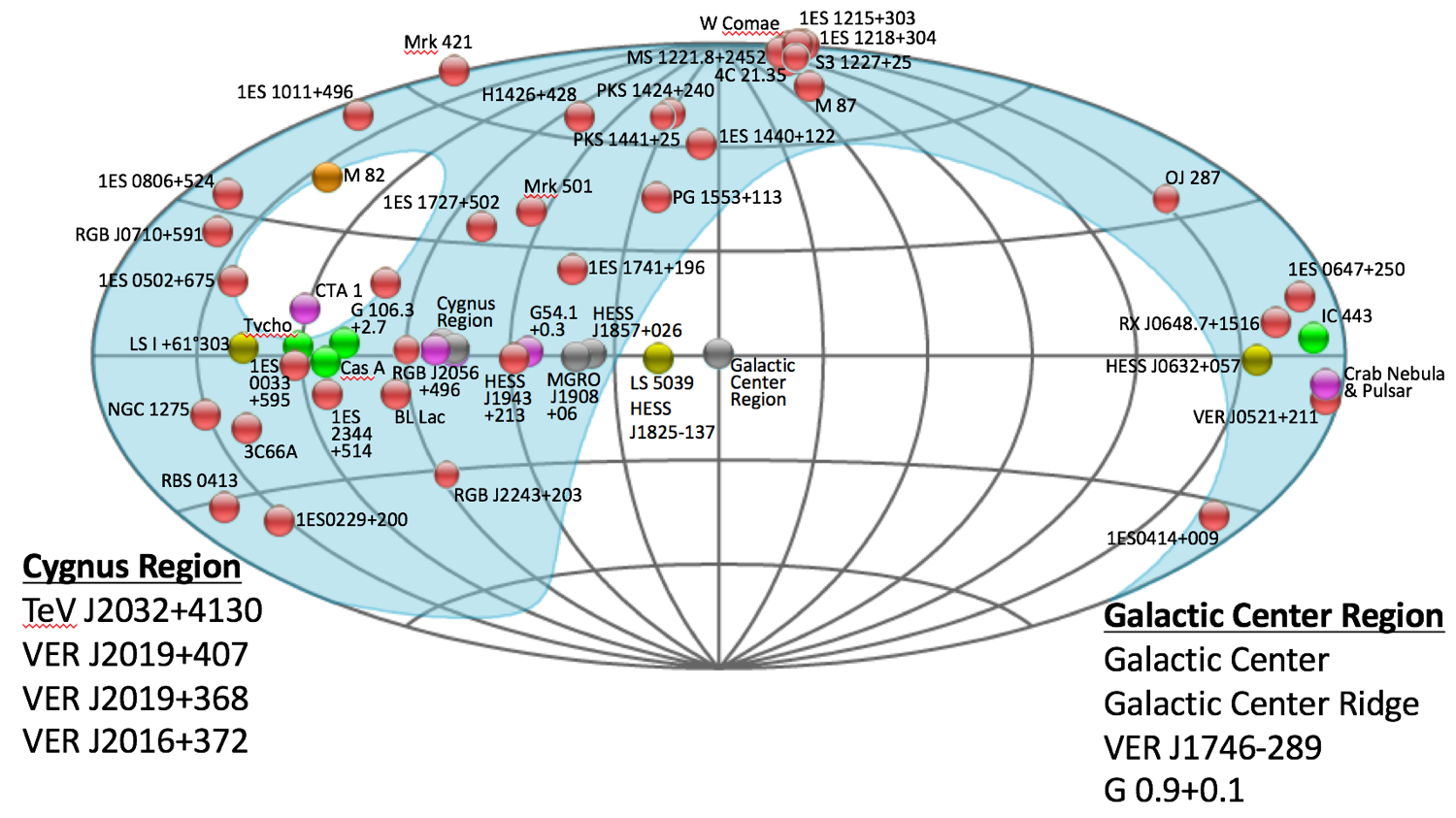} } }
   \caption{{\footnotesize The VERITAS sky catalog in Galactic
       coordinates. Different astrophysical classes are shown with
       different colored markers.  The red circles are AGN.  The blue
       region is the sky visible to VERITAS in normal observing modes.}}
   \label{VERITAS_catalog}
\vspace{-0.3cm}
 \end{figure*}

\section{VERITAS AGN Program}

VERITAS \cite{VERITAS_spec} is located at the F.L. Whipple Observatory in
southern Arizona, USA  (31$^{\circ}$ 40' N, 110$^{\circ}$ 57' W,  1.3 km a.s.l.).  It is most
sensitive between $\sim$85 GeV and $\sim$30 TeV and the full array has been used to
regularly observe the Northern sky each season (September $-$ July) since 
2007.  Following the completion of a series of upgrades in Summer 2012, the observatory
can detect an object with flux equal to 1\% Crab Nebula flux (1\%
Crab) in $\sim$25 hours, and spectral reconstruction of observed signals can be performed above
$\sim$100 GeV.  The typical systematic errors reported for scientific
quantities are $\sim$20\% on the flux and 0.1 on the photon index ($\Gamma$).

Over the past decade, VERITAS has acquired an average of $\sim$950 h
of good-weather observations each season during ``dark time'' (moon
illumination $<$30\%).   The collaboration also has the capability to observe during periods of
``bright moonlight'' (i.e. $>$30\% illumination) \cite{BrightMoon_paper}.  One of the
bright-moon techniques has comparable sensitivity to
the normal observing mode and only slightly higher threshold (e.g. 250 GeV).
Since 2012, it is a regular component of VERITAS data taking, adding another 
$\sim$30\% to overall good-weather data yield.  VERITAS has
successfully detected several blazars during these observations (see, e.g., \cite{VERITAS_1727}).

AGN observations are a significant component of VERITAS data taking
($\sim$50\% since 2007).  As of June 2017, they comprise a total of $\sim$4240h ($\sim$425 h per
year) of good-weather dark time and $\sim$840h ($\sim$170 h per year) of
good-weather, bright-moon time.  The dark time is split  $\sim$90\% to blazars, primarily BL Lac objects, 
and $\sim$10\% to radio-galaxies, primarily M\,87.  The bright-moon
time is almost entirely observations of BL\,Lac objects, with
observationssplit $\sim$35\% to hard-spectrum VHE blazars and
$\sim$65\% to candidates for new VHE discoveries during the past two
seasons.   AGN comprise 62\% of the VERITAS source catalog (shown in Figure~\ref{VERITAS_catalog}),
 and a table listing 34 of the 36 AGN detected by VERITAS can be found
in \cite{Benbow_ICRC2015}; the two newest detections (RGB\,J2056+496
and OJ\,287) are described below.

Conceptually, much of the VERITAS AGN program is based on regular
($\sim$weekly) monitoring observations of the entire Northern VHE
catalog when the targets are best visible to self-identify relatively long-term VHE flaring 
episodes.  These monitoring observations are supplemented with
coordinated data at lower energy, so that long-term
contemporaneous MWL data sets exist for all Northern VHE AGN.
When interesting flaring events are observed in any of these
monitoring programs, intense MWL target-of-opportunity (ToO)
observations are immediately scheduled.  The monitoring program is
designed such that the minimum cadence will detect a 10\% Crab
flux.  Over a period of many years, even the lowest level monitoring significantly increases the
data set for each object, and for $\sim$10 particularly interesting
targets longer-duration monitoring is scheduled to generate very deep,
legacy exposures.

While the philosophy of the AGN program is largely based on exploiting the existing catalog via deep / timely measurements of
the known sources, $\sim$30\% of the AGN program was devoted to the discovery and
follow-up observations of new VHE AGN during the past two seasons.
Most discovery observations are taken on targets from a list of
selected candidates during bright-moon time, or ToO observations
triggered by MWL partners.  Recently, our selected discovery
candidates have focused on a comprehensive list of Northern objects
including all the hardest ($\Gamma_{2FHL} < 2.8$) AGN in the {\it Fermi}-LAT 2FHL ($>$50 GeV) catalog \cite{2FHL_Catalog}, and
the X-ray brightest HBLs in the 2WHSP catalog (i.e. objects with a
``TeV Figure of Merit'' $>$ 1.0) ~\cite{2WHSP_cat}. 

Regardless of whether their focus is the discovery and/or follow-up
observation of a new VHE source, or exploiting the potential of bright
flares in known VHE blazars, ToO observations are a key component of the VERITAS AGN 
program.  These data average  $\sim$25\% of the AGN yield each season.
Indeed it is notable that aside from some low-level monitoring, nearly all VERITAS
FSRQ data are taken via ToO observations.

\section{Recent Highlights}

\begin{figure*}[!t]
   \centerline{
     {\includegraphics[width=2.5in]{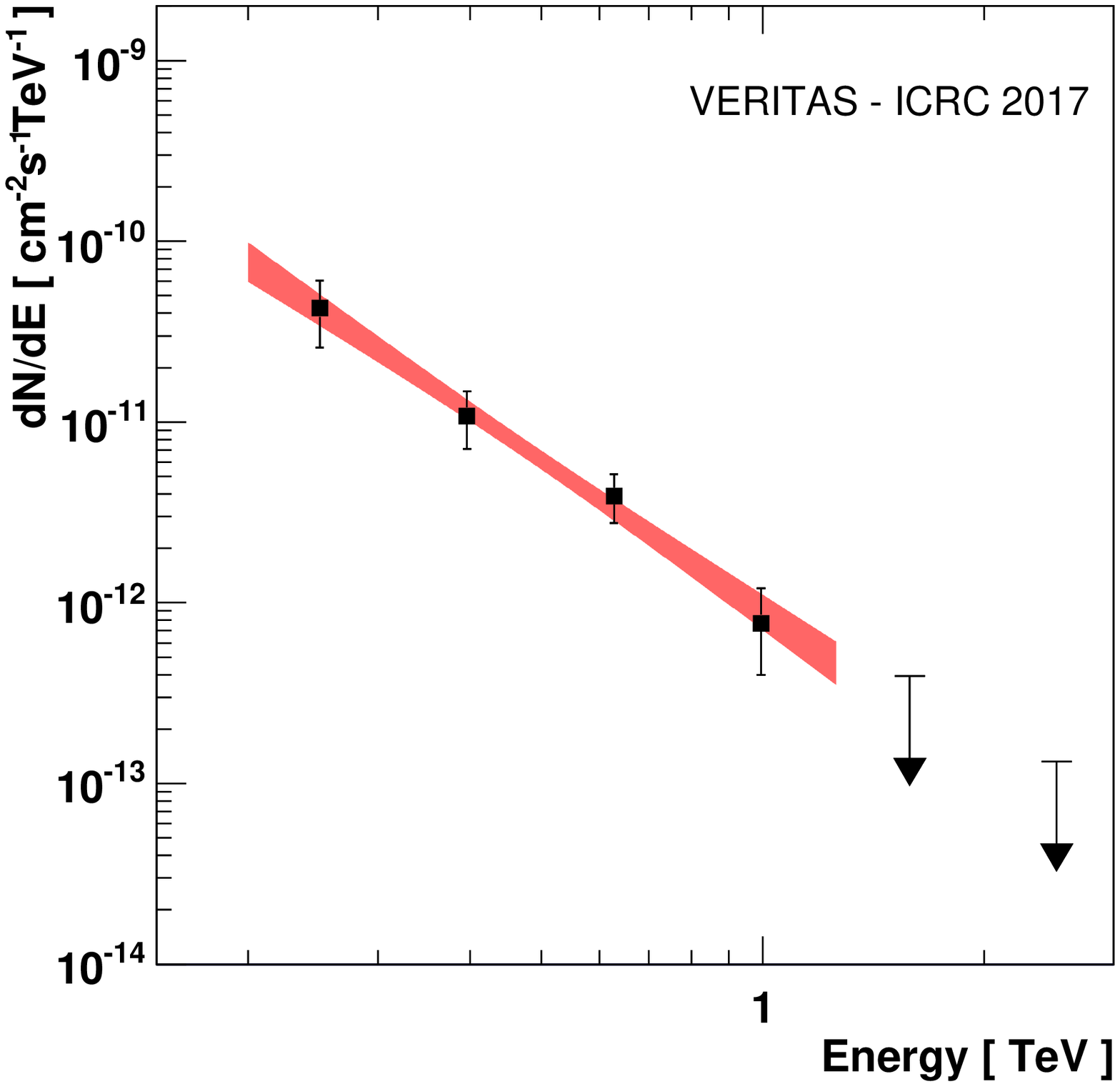} }
     \hfil
     {\includegraphics[width=2.5in]{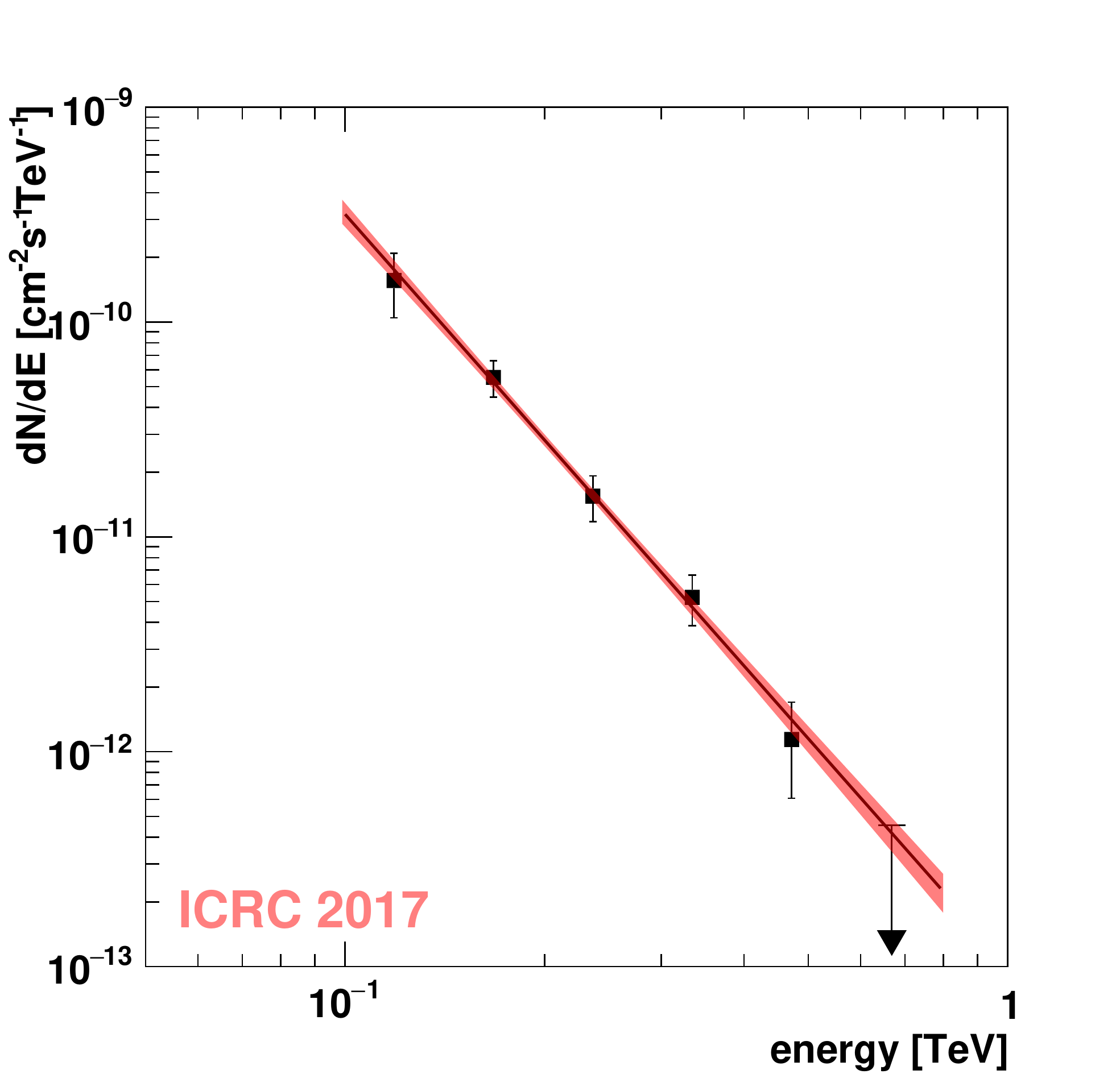} }
   }
   \caption{{\footnotesize Left) The preliminary VHE spectrum measured from
     the direction of RGB\,J2056+496.  The best fit ($\chi^2$ = 0.46,
     NDF = 2) of a power law $dN/dE = I_o \times (E / E_o)^{-\Gamma}$
     to these data has photon index $\Gamma = 2.77 \pm 0.40_{\rm stat}
     \pm 0.10_{\rm syst}$, $I_o = (1.15 \pm 0.23_{\rm stat} \pm  0.23_{\rm syst}) \times 10^{-11}$ cm$^{-2}$ s$^{-1}$ TeV$^{-1}$, and $E_o$ = 400 GeV.  
     Right)   The preliminary VHE spectrum measured from 
     OJ\,287 \cite{OJ287_ICRC}. The best fit of a power law yields photon index $\Gamma = 3.49 \pm 0.28_{\rm stat} \pm 0.10_{\rm syst}$.}}
   \label{Misc_spec}
\vspace{-0.2cm}
 \end{figure*}

\begin{figure*}[!t]
   \centerline{
     {\includegraphics[width=3.0in]{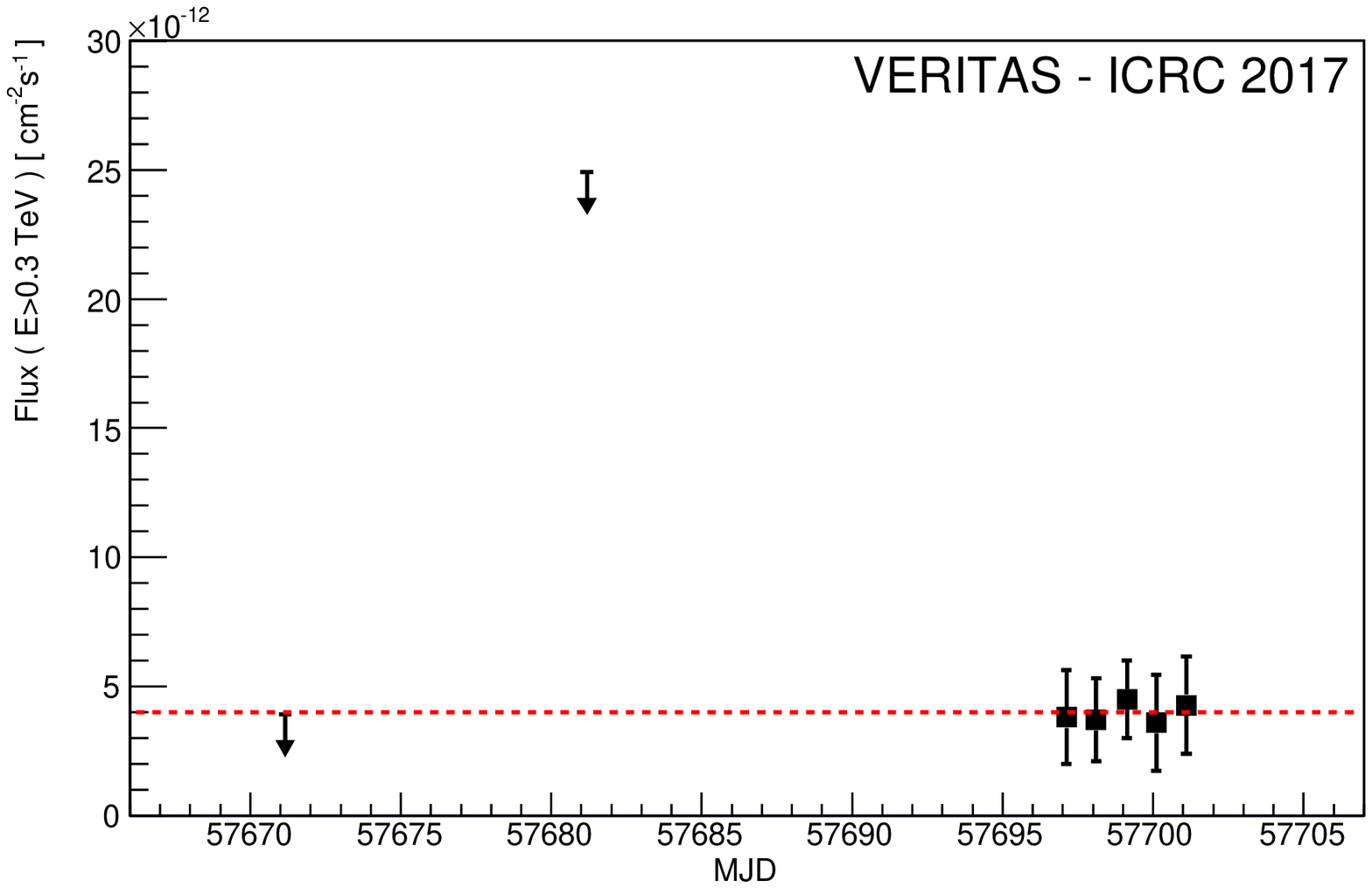} }
     \hfil
     {\includegraphics[width=2.7in]{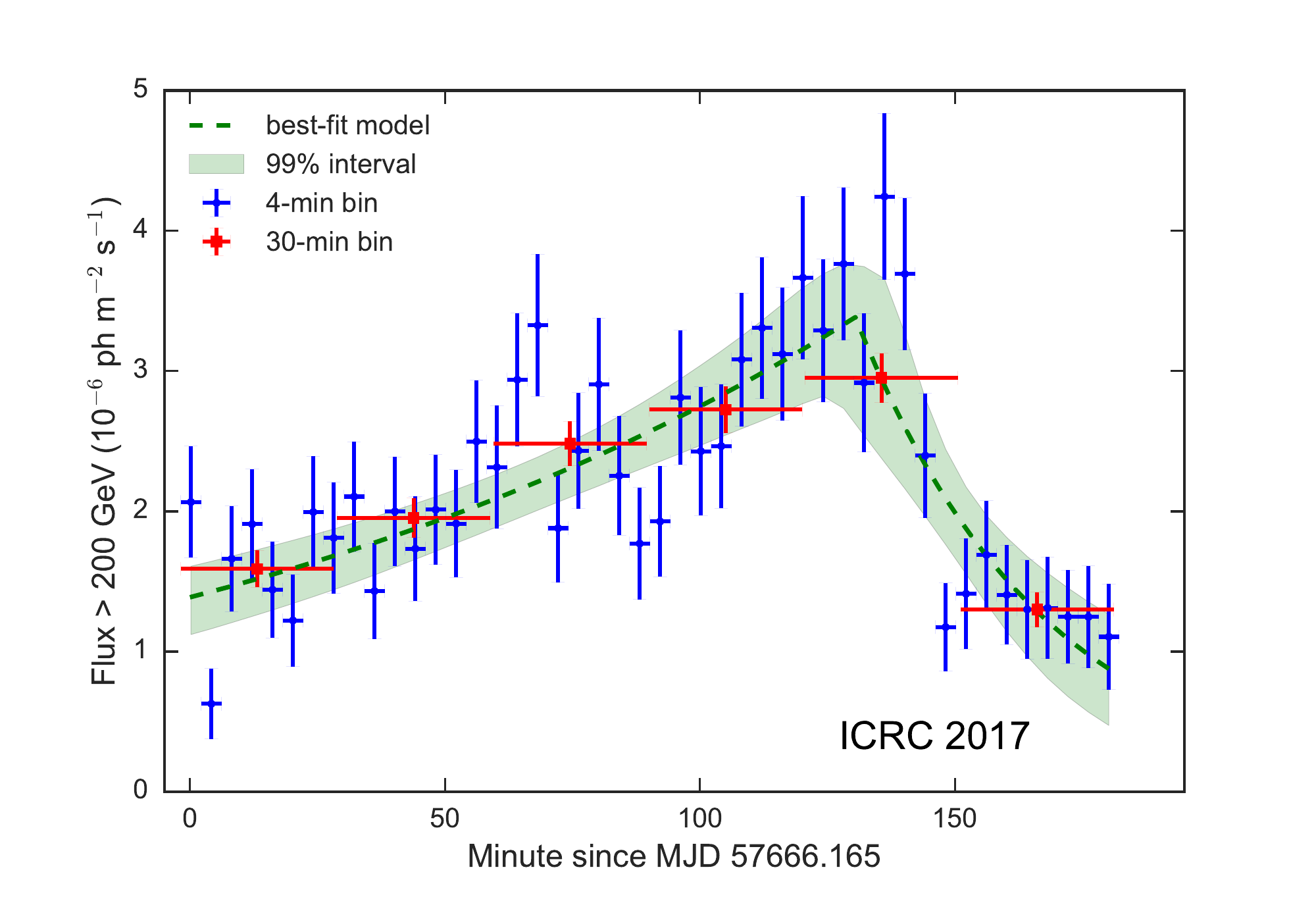} } 
   }
   \caption{{\footnotesize Left) The preliminary VHE light curve
       measured from the position of RGB\,J2056+496 in 2016. The
       dashed line is the best fit of a constant to the data.
       Right) The preliminary VHE light curve measured from BL\,Lac on
       October 5, 2016, in 4-minute (blue dots) and 30-minute (red
       squares) bins.  The dashed line and shaded region show the best
     fit mode and its 99\% confidence interval, respectively.}}
   \label{Misc_LCs}
\vspace{-0.2cm}
 \end{figure*}

{\bf RGB\,J2056+496} is one of the brightest, hard-spectrum ($\Gamma_{2FHL} \sim 2.3$) objects in the
{\it Fermi}-LAT 2FHL ($>$50 GeV) catalog \cite{2FHL_Catalog} not directly observed
prior to the 2016-17 season by VERITAS.  This blazar, of unknown
redshift, is also a XMM-Newton source and a Swift-BAT source.
It was observed with the array for 9.0 h of good-quality live time between October 10, 2016 and November 9, 2016.
A preliminary analysis of these observations yields an excess of 120 events above the background at the position of the blazar, 
corresponding to a statistical significance of 6.3 standard
deviations ($\sigma$).   The VERITAS collaboration is interpreting this excess as the 
discovery of VHE $\gamma$-ray emission from the blazar, noting that
the blazar is only $\sim$13 arc-seconds from the micro-quasar candidate LS\,III\,+49\,13, and the two objects are
effectively co-located for VERITAS.  The observed VHE
spectrum from RGB\,J2056+496 is shown in Figure~\ref{Misc_spec}.
The observed integral flux is F($>$300 GeV) = ($4.00 \pm
0.77_{\rm stat} \pm 0.80_{\rm syst}) \times 10^{-12}$ cm$^{-2}$ s$^{-1}$, 
or about 2.9\% Crab above the same threshold.
The nightly VHE light curve shown in Figure~\ref{Misc_LCs}, and the
flux in November 2016 is well fit by a constant ($\chi^2$ = 0.22, NDF = 4).  Four Swift XRT / UVOT observations were
taken simultaneous with the November VERITAS observations of this blazar, which 
comprise $\sim$85\% of the data set.  While SED
modeling will be possible, we note the brightness of LS\,III\,+49\,13
(V $\sim$ 8.8) makes optical photometry and
spectroscopy of the AGN very difficult.

{\bf OJ\,287} is an optically-bright blazar at redshift $z =
0.306$.  It has unusual optical behavior, displaying regular outbursts with a overall period of
$\sim$12 years, but also significant divergences from this period.
This quasi-periodicity has often been explained and predicted by the
presence of a binary black hole system at the core of OJ\,287.  Its SED has
shown HBL-like features suggesting it is an appropriate target for VHE
observatories \cite{Costamante}.  The extrapolation of its {\it Fermi}-LAT 3FGL
spectrum ($\Gamma_{3FGL} \sim $ 2.1) to the VHE band
is also favorable ($\sim$10\% Crab), but OJ\,287 is not in the 2FHL
($>$50 GeV) catalog \cite{2FHL_Catalog}.   In December 2007, during an anticipated phase
of optical brightness, VERITAS observed OJ\,287 for
$\sim$10 h live time, yielding an upper limit of $\sim$2.6\% Crab
above 180 GeV \cite{VERITAS_Blazar_UL}.  Following the observation of
long-lasting, enhanced X-ray activity from OJ\,287 by Swift in
late-2016, VERITAS began a series of ToO observations of the blazar.
In February 2017, these observations resulted in the successful
discovery of VHE emission from OJ\,287 (ATel \#10051) which coincided with the
brightest state yet observed with the Swift XRT (ATel \#10043).  The VHE
discovery and bright X-ray state led to an intense VERITAS and 
MWL observation campaign though March 2017.  A total of $\sim$50 h of
quality-selected live time were acquired on OJ\,287 with the VERITAS
array.  During these observations a point-like excess of 556
$\gamma$-rays is detected from OJ\,287, corresponding to a statistical
significance of 9.7$\sigma$.  The VHE flux observed
by VERITAS is variable, with a time-average of F($>$150 GeV) = ($4.61 \pm
0.61_{\rm stat} \pm 0.92_{\rm syst}) \times 10^{-12}$ cm$^{-2}$
s$^{-1}$, or about 1.3\% Crab above the same
threshold.  The observed VHE spectrum is shown in
Figure~\ref{Misc_spec}.
An early look at the copious MWL data from this campaign suggests a
shift of the high-frequency SED peak to higher energies during the VHE detection.
More details on the VERITAS and MWL observations can be found in these
proceedings \cite{OJ287_ICRC}.  

{\bf BL\,Lacertae} ($z = 0.069$) was initially discovered as a weak (3\% Crab) VHE emitter
during a flare in August 2005 \cite{MAGIC_BLLac}.  Since 2010, it has been regularly monitored with the
VERITAS array yielding a total exposure of nearly 70 h of
good-weather data including ToO observations. This long-term
monitoring has shown BL\,Lacertae is not normally detectable in the
VHE band, however, there have been 4 flares detected by VERITAS.
The first was a brief flare in June 2011, that peaked at a
VHE flux of $\sim$125\% Crab \cite{BLLac_paper}.  This flare had a  rapid exponential decay
($\tau = 13 \pm 4$ min) and was associated with the appearance 
of a superluminal radio knot .  Two smaller flares of
$\sim$16\% Crab and $\sim$9\% Crab were recorded on the nights
of June 21, 2015 and November 30, 2015, respectively
\cite{BLLac2_ICRC}.    An exceptional flare was detected with VERITAS
during routine monitoring on October 5, 2016, triggering immediate ToO observations.  A total
exposure of 2.6 h good-quality live time was acquired, yielding a strong
detection ($\sim$71$\sigma$).  The light curve from this event is shown
in Figure~\ref{Misc_LCs}.   The flare peaks at $\sim$180\% Crab when binned
on 4-minute time scales (or $\sim$125\% Crab when binned on 30-minute
scales) and it is characterized by a slow rise (t $\sim$ 140$^{+25}_{-11}$
min) followed by a more rapid fall (t $\sim$ 36$^{+8}_{-7}$ min).
Similar to the 2011 event, this VHE flare is also associated with the appearance of a candidate
superluminal radio knot.  More details regarding the VERITAS flare as
well as interpretation of contemporaneous VLBA, optical (photometry and
polarization), Swift, and {\it Fermi}-LAT data can be found in these
proceedings \cite{BLLac_ICRC}.

{\bf 1ES\,1959+650} is a well-known, nearby ($z = 0.047$) HBL that has been
relatively inactive since a possible "orphan flare" in 2002.  During the
2015-16 season, the X-ray and MeV-GeV fluxes were the highest seen from this blazar since the start of the
{\it Swift} and {\it Fermi}-LAT missions.  VHE flaring was also observed by
VERITAS during deep observations in both Fall 2015, and in Spring 2016
 ($>$15 h each, $>$80$\sigma$ detections).  The peak flux observed during each of 
the events was $\sim$200\% Crab and $\sim$300\% Crab, respectively.
More details on the significant MWL coverage and the VERITAS
observations can be found in these proceedings \cite{1ES1959_ICRC}.

\begin{figure*}[!t]
   \centerline{ {\includegraphics[width=5.5in]{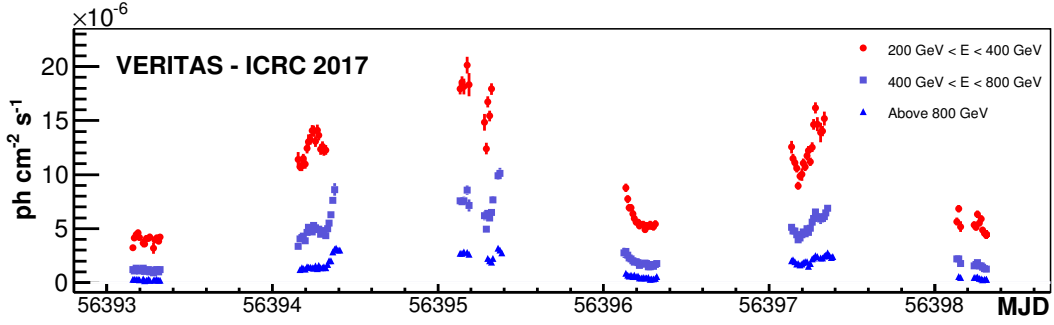} } }
   \caption{{\footnotesize The light curve measured from Mrk 421
       in April 2013.  The integral fluxes are shown for three separate energy
       ranges: 200 GeV $<$ E $<$ 400 GeV, 400 GeV $<$ E $<$ 800 GeV, and
       E $>$ 800 GeV. The fluxes are provided at the VERITAS public web
       page: https://veritas.sao.arizona.edu/. }}
   \label{Mkn421_2013_fig}
\vspace{-0.2cm}
 \end{figure*}

{\bf Mrk\,421} ($z=0.030$) is the most well-studied VHE blazar and VERITAS
has acquired more than 300 h of good-weather data on this object.
These studies include data taken during both low-states and
exceptionally bright states, and nearly all data have been taken
contemporaneously with a wide assortment of MWL partners.  In 2013, VERITAS embarked on
a deep, season-long MWL observing campaign on Mrk 421 that was unprecedented 
due to its hard X-ray coverage from the NuStar satellite.
While relatively low fluxes were recorded by VERITAS and other TeV partners for
from January to March 2013 \cite{Mrk421_NuStar}, exceptional flaring activity
was recorded by VERITAS and MAGIC in April 2013 (ATel \#4976).  This led to $\sim$12 h / night of uninterrupted
VHE coverage with MAGIC and VERITAS, as well as near
continuous NuStar coverage, pointed {\it Fermi}-LAT observations, numerous
Swift pointings, and copious coverage at optical and longer
wavelengths for several nights.  A total of 21.8 h of good-quality
live time were acquired by VERITAS between April 11 and April 16, 2013.
The VERITAS light curve from these
observations is shown in Figure~\ref{Mkn421_2013_fig}.  A summary of
results from the data is given in Table~\ref{Mkn421_table}.

\begin{table}[t]
\begin{center}
\begin{tabular}{c | c | c | c | c | c}
\hline
{\footnotesize Date} & {\footnotesize  Live Time} &  {\footnotesize
  F($>$280 GeV)} &  {\footnotesize  F($>$280 GeV)} & {\footnotesize
  $\Gamma$} & {\footnotesize $E_{cut}$} \\
 & {\footnotesize [ h ]} & {\footnotesize [ cm$^{-2}$ s$^{-1}$ ] } & 
 {\footnotesize [ Crab ]} &  & {\footnotesize [ TeV ]} \\
\hline
\vspace{-0.13cm}
{\footnotesize April 11} & {\footnotesize  2.7} & {\footnotesize
  $(2.67 \pm 0.07) \times 10^{-10}$} & {\footnotesize  1.87} &
{\footnotesize $2.09 \pm 0.11$} & {\footnotesize $0.81 \pm 0.13$}  \\
\vspace{-0.13cm}
{\footnotesize April 12} & {\footnotesize  5.0} & {\footnotesize
  $(8.16 \pm 0.09) \times 10^{-10}$} & {\footnotesize  5.70} & {\footnotesize $2.04 \pm 0.03$} & {\footnotesize$2.27 \pm 0.14$} \\
\vspace{-0.13cm}
{\footnotesize April 13} &{\footnotesize 2.1} & {\footnotesize 
$(10.83  \pm 0.14) \times 10^{-10}$} & {\footnotesize  7.57} & {\footnotesize $2.03 \pm 0.05$} & {\footnotesize$1.96 \pm 0.17$} \\
\vspace{-0.13cm}
{\footnotesize April 14} & {\footnotesize 3.9} &{\footnotesize 
$(4.04  \pm 0.07) \times 10^{-10}$} & {\footnotesize  2.82} & {\footnotesize $2.18 \pm 0.06$} & {\footnotesize$1.24 \pm 0.16$} \\
\vspace{-0.13cm}
{\footnotesize April 15} & {\footnotesize 5.1} & {\footnotesize 
$(8.86  \pm 0.09) \times 10^{-10}$} & {\footnotesize  6.19} & {\footnotesize $2.06 \pm 0.02$} &{\footnotesize $4.58 \pm 0.34$} \\
{\footnotesize April 16} & {\footnotesize 2.0} & {\footnotesize 
$(3.65  \pm 0.09) \times 10^{-10}$} & {\footnotesize  2.55} & {\footnotesize $2.44 \pm 0.11$} & {\footnotesize$1.75 \pm 0.46$} \\
\hline
\end{tabular}
\vspace{-0.2cm}
\caption{{\footnotesize VERITAS results from Mrk\,421 in April 2013.
    For each night's observation, the average integral
    flux above 280 GeV, and key parameters from the best fit of a power law with an
    exponential cutoff to the spectra are shown.}}
\label{Mkn421_table}
\end{center}
\vspace{-0.9cm}
\end{table}

Over the past two seasons VERITAS has devoted approximately 90 hours
toward observing radio galaxies. A particular highlight of these observations
is the detection of two bright flares (ATel \#9690 and ATel \#9931)
from {\bf NGC 1275}. The first flare on October 30, 2016, was
$\sim$15\% Crab flux and the second flare on January 2, 2017, was $\sim$65\%
Crab flux.  Both detections were at least an order of magnitude brighter
than previous flare detections ($\sim$1\% Crab; see, e.g.,
\cite{Benbow_ICRC2015}). Another recent highlight is our
participation in an international MWL campaign on M\,87 in March-April
2017.  Here we coordinated our observations with the new Event Horizon
Telescope to test if any high-energy flickering in the low-emission
state is related with any new phenomena near the core of M\,87.
The VHE flux during our $\sim$25 h of observations
was consistent with the low state and our total exposure on M\,87 is now $\sim$290 h.
Although it had been nearly five years since any VERITAS observations were taken with the goal of discovering
new radio galaxies, $\sim$15 h exposures were taken in 2016-17 on each
of 3C\,303  and 3C\,264.  No significant excesses were found in these data.

\section{Conclusion}

VERITAS has now acquired more than 10,000 hours of good-weather, scientific
observations, including more than 5,000 hours targeted on AGN.  
The array continues to run well with $>$1250 h of observations 
acquired in each of the two most recent seasons, including $>$1100 h of
good-weather data taken on AGN since {\it ICRC2015}.   These most
recent AGN observations have resulted in the VHE discoveries of 2 BL Lac objects,
and several interesting VHE flares.  The VERITAS AGN catalog now
includes 32 BL Lac objects, 
2 FSRQs and 2 FR I radio galaxies.  The VERITAS collaboration plans to operate the
telescope array through at least 2019, and has correspondingly organized a strategy to guide
the AGN program through this time (see \cite{Benbow_ICRC2015} for more
details).  This program is heavily focused on
regular VHE and MWL monitoring of all known VHE AGN in the Northern
Hemisphere, and emphasizes immediate and intense ToO follow-up of interesting
flaring events.  Although the VERITAS collaboration has published
upper limits on more than 100 AGN \cite{VERITAS_Blazar_UL}, and has observed nearly 100 other
AGN since, we maintain a strong program focused on the discovery of new VHE
blazars.  Here we are using ToO observations of
interesting VHE candidates during dark time and non-ToO observations of
selected targets with limited existing exposures during
bright-moon time.  As AGN observations remain a priority for
the VERITAS collaboration, we expect our long tradition of
producing exciting results to continue.

\vspace{0.1cm}

\footnotesize{This research is supported by grants from the U.S. Department of
Energy Office of Science, the U.S. National Science Foundation and the
Smithsonian Institution, and by NSERC in Canada. We acknowledge the
excellent work of the technical support staff at the Fred Lawrence
Whipple Observatory and at the collaborating institutions in the
construction and operation of the instrument.  The VERITAS
Collaboration is grateful to Trevor Weekes for his seminal
contributions and leadership in the field of VHE gamma-ray
astrophysics, which made this study possible.}

\end{document}